\documentclass[12pt,preprint]{aastex}

\received{7 February 2002}
\revised{1 May 2002}
\accepted{7 May 2002}

\cpright{AAS}{2002}

\shortauthors{Hjorth et al.}
\shorttitle{GRB 980613: Afterglow and environment}

\slugcomment{Submitted to ApJ, 7 February 2002; accepted 7 May 2002}

\newcommand{\etal}{et~al.}

\newcommand{\hst}{{\sl HST\/}}
\newcommand{\sax}{{\sl BeppoSAX\/}}

\begin{document}

\title{The afterglow and complex environment of the optically dim 
       burst \objectname{GRB~980613}\footnote{\rm Based on observations 
		 with the Nordic
                 Optical Telescope, which is operated on the island of
                 La Palma jointly by Denmark, Finland, Iceland,
                 Norway, and Sweden, at the Spanish Observatorio del
                 Rocque de los Muchachos of the Instituto de
                 Astrofisica de Canarias, and on observations with the
                 NASA/ESA {\sl Hubble Space Telescope}, obtained at
                 the Space Telescope Science Institute, which is
                 operated by the Association of Universities for
                 Research in Astronomy, Inc.\ under NASA contract
                 NAS5-26555.}}

\author{Jens Hjorth\altaffilmark{2,3,4},     % jens@astro.ku.dk
Bjarne Thomsen\altaffilmark{5},              % bt@ifa.au.dk
Svend R. Nielsen\altaffilmark{5},            % runge@ifa.au.dk
Michael I. Andersen\altaffilmark{6,7},       % manderse@sun3.oulu.fi
Stephen T. Holland\altaffilmark{5,8},        % sholland@nd.edu
Johan U. Fynbo\altaffilmark{9},              % jfynbo@eso.org
Holger Pedersen\altaffilmark{2},             % holger@astro.ku.dk
Andreas O. Jaunsen\altaffilmark{10,3,11},    % ajaunsen@eso.org
Jules P. Halpern\altaffilmark{12},           % jules@astro.columbia.edu
Robert Fesen\altaffilmark{13},               % fesen@snr.dartmouth.edu
Javier Gorosabel\altaffilmark{14,15,16},     % jgu@dsri.dk
Alberto Castro-Tirado\altaffilmark{15,16},   % ajct@iaa.es + laeff
Richard G. McMahon\altaffilmark{17},         % rgm@ast.cam.ac.uk
Michael D. Hoenig\altaffilmark{17},          % mhoenig@ast.cam.ac.uk
Gunnlaugur Bj\"ornsson\altaffilmark{18},     % gulli@raunvis.hi.is
Lorenzo Amati\altaffilmark{19},              % amati@tesre.bo.cnr.it
Nial R. Tanvir\altaffilmark{20},             % nrt@star.herts.ac.uk
and
Priyamvada Natarajan\altaffilmark{21}}       % priya@astro.yale.edu

\altaffiltext{2}{Astronomical Observatory, University of Copenhagen,
Juliane Maries Vej 30, DK--2100 Copenhagen~\O, Denmark; 
\email{jens@astro.ku.dk}}%
\altaffiltext{3}{Center for Advanced Study, Drammensveien 78, Oslo, Norway}
\altaffiltext{4}{NORDITA, Blegdamsvej 17, DK--2100 Copenhagen~\O, Denmark}
\altaffiltext{5}{Institute of Physics and Astronomy, University of Aarhus,
DK--8000 {\AA}rhus C, Denmark}
\altaffiltext{6}{Division of Astronomy, University of Oulu, P. O. Box 3000,
FIN--90014 Oulu, Finland} 
\altaffiltext{7}{Nordic Optical Telescope, Apartado 474, E--38700 
St.~Cruz de La Palma, Canary Islands, Spain}
\altaffiltext{8}{Department of Physics, University of Notre Dame, Notre Dame,
IN 46556--5670, USA}
\altaffiltext{9}{European Southern Observatory, Karl-Schwarzschild-Strasse 2,
D-85748 Garching bei M{\"u}nchen, Germany}
\altaffiltext{10}{Institute of Theoretical Astrophysics, University of Oslo,
Pb.~1029 Blindern, N--0315~Oslo, Norway}
\altaffiltext{11}{European Southern Observatory, Casilla 19001, Santiago 19,
Chile}
\altaffiltext{12}{Columbia Astrophysics Laboratory, Columbia University,
550 West 120th Street, New York, NY 10027, USA}
\altaffiltext{13}{Department of Physics and Astronomy, Dartmouth College,
6127 Wilder Laboratory, Hanover, NH 03755--3528, USA}
\altaffiltext{14}{Danish Space Space Research Institute,
Juliane Maries Vej 30, DK--2100 Copenhagen~\O, Denmark}
\altaffiltext{15}{Instituto de Astrof\'\i sica de Andaluc\'\i a, IAA-CSIC,
Granada, Spain}
\altaffiltext{16}{Laboratorio de Astrof\'{\i}sica Espacial y F\'{\i}sica 
Fundamental (LAEFF--INTA), P. O. Box 50.727, E--28080 Madrid, Spain}
\altaffiltext{17}{Institute of Astronomy, Madingley Road, Cambridge CB3 0HA, UK}
\altaffiltext{18}{Science Institute, University of Iceland, Dunhagi 3, 
IS--107 Reykjavik, Iceland}
\altaffiltext{19}{Instituto Tecnologie e Studio Radiazoni Extraterresti, CNR,
Via Gobetti 101, 40129 Bologna, Italy}
\altaffiltext{20}{Department of Physical Sciences, University of Hertfordshire, 
College Lane, Hatfield, Hertfordshire AL10 9AB, UK}
\altaffiltext{21}{Department of Astronomy, Yale University, 265 Whitney Avenue,
New Haven, CT 06511, USA}

\begin{abstract}

     We report the identification of the optical afterglow of 
\objectname{GRB 980613} in $R$- and $I$-band images obtained between 
16 and 48 hours after the gamma-ray burst. Early near-infrared (NIR) $H$ and 
$K^\prime$ observations are also reported. The afterglow was optically faint 
($R\approx 23$) at discovery but did not exhibit an unusually rapid decay 
(power-law decay slope $\alpha < 1.8$ at 2$\sigma$). The optical/NIR 
spectral index ($\beta_{\rm RH} < 1.1$) was consistent with the 
optical-to-$X$-ray spectral index ($\beta_{\rm RX} \approx 0.6$), indicating 
a maximal reddening of the afterglow of $\approx 0.45$ mag in $R$.  Hence
the dimness of the optical afterglow was mainly due to the fairly flat
spectral shape rather than internal reddening in the host galaxy.  We also 
present late-time {\sl HST\/}/STIS images of the field in which 
\objectname{GRB 980613} occurred, obtained 799 days after the burst. These 
images show that \objectname{GRB 980613} was located close to a very compact, 
blue $V=26.1$ object inside a complex region consisting of star-forming 
knots and/or interacting galaxy fragments. Therefore, \objectname{GRB 980613} 
constitutes a strong case for the association of cosmological gamma-ray bursts 
with star-forming regions.

\end{abstract}

\keywords{
cosmology: observations ---
gamma rays: bursts ---
galaxies: star formation  ---
galaxies: photometry 
}

\section{Introduction}

     Gamma-ray bursts (GRBs) are the most energetic events in the Universe. 
The currently favored models for long-duration GRBs (lasting more than about 
one second) involve the collapse of a very massive progenitor star (see e.g., 
\citet{macfadyen99}, \cite{vietri98}, and \citet{bloom00}).  Because of the 
short-lived nature of massive stars, GRBs are believed to originate in regions 
where stars are born.  Hence, in this scenario, GRBs are expected to be
efficient tracers of massive-star formation in galaxies 
\citep{paczynski98}.  Consequently, GRBs may trace the (massive) 
star-formation rate in the Universe as a function of redshift 
\citep{totani97,wijers98,blain00}.  
Evidence supporting this hypothesis includes (i) the association of 
\objectname{SN1998bw} and \objectname{GRB 980425} which occurred in a 
star-forming region \citep{galama98,fynbo00a}, (ii) the association of 
cosmological GRBs with supernovae 
(e.g., Bloom et al.\ 1999; Galama et al.\ 2000; Bj\"ornsson et al.\ 2001;
Bloom et al.\ 2002; Garnavich et al.\ 2002),
%\citep{bloom99,galama00,bjornsson01}, 
(iii) the discovery of 
$X$-ray absorption edges and lines in GRB afterglows \citep{amati00,piro00,
reeves02}, (iv)
the locations of GRBs close to star-forming regions in interacting galaxies
\citep{holland99}, in starburst or ultraluminous
infrared galaxies (ULIRGs)
\citep{djorgovski98a,holland00a,berger01} or in Ly$\alpha$-emitting
galaxies \citep{fynbo02}, (v) indications of circumburst density
distributions similar to those generated by stellar winds \citep{jaunsen00}, 
(vi) the existence of GRBs at very high redshifts \citep{andersen00}, and 
(vii) the inferred high column densities of neutral Hydrogen along the line 
of sight to GRBs \citep{wijers01}.

     \objectname{GRB 980613} triggered the \sax\ Gamma-Ray Burst
Monitor (40--700 keV, \citet{costa98a}) on 1998 June 13 at 04:51:06 UT
and was simultaneously detected and positioned at the equatorial
coordinates R.A.~(J2000.0) = 10h17m44s, Dec.~(J2000.0) =
+71$^\circ$29\farcm9 (4\arcmin\ error radius) by the Wide Field
Cameras instrument (2--26 keV, \citet{jager97}) on board the same
satellite \citep{smith98}.  The event was also detected by BATSE/{\sl CGRO\/}
with a flux and fluence near the middle of the BATSE burst distribution
\citep{woods98}. As measured by \sax, it was composed of a single pulse
of duration $\approx 15$~s in $\gamma$-rays and $\approx 50$~s in
$X$-rays. The peak flux and fluence were $(1.3\pm 0.4)\times 10^{-7}$
erg~cm$^{-2}$~s$^{-1}$ and $(1.0\pm 0.2)\times 10^{-6}$ erg~cm$^{-2}$
in the 40--700 keV energy band, and $(2.0\pm 0.3)\times 10^{-8}$
erg~cm$^{-2}$~s$^{-1}$ and $(2.7\pm 0.4)\times 10^{-7}$ erg~cm$^{-2}$
in the 2--26~keV energy band.  The \sax\ Narrow Field Instruments
\citep{boella97} were pointed to the \objectname{GRB 980613} error box
about 9~h after the burst for a total time of about 100~ks, leading to
the discovery of a possible $X$-ray afterglow source at R.A.~(J2000.0)
= 10h17m53s, Dec.~(J2000.0) = +71$^\circ$27\arcmin24\arcsec\
(1\arcmin\ error radius) with an average flux of $(1.1\pm 0.3)\times
10^{-13}$ erg~cm$^{-2}$~s$^{-1}$ \citep{costa98b}.  \objectname{GRB
980613} is the faintest burst detected by \sax\ in the $\gamma$-ray
band but is relatively rich in $X$-rays \citep{costa98c,costa99}. 
It belongs to the less luminous $X$-ray class of \citet{boer00}.

     Following the discovery of the optical afterglow by
\citet{hjorth98} (described in more detail in this paper) an object
identified as the host galaxy was detected by \citet{djorgovski98b}
and suggested to be part of an interacting system by
\citet{hjorth99}. The complex region in which \objectname{GRB 980613}
occurred was studied in more detail by \citet{djorgovski00} who
concluded that the burst was located in a starburst galaxy at
$z=1.0969\pm0.0002$ which is interacting with one or more of several
galaxies or galaxy fragments in the vicinity of the GRB. This led
\citet{djorgovski00} to propose that the GRB originated from a
merger-induced starburst.

     We here present the initial identification and
optical/near-infrared follow-up observations of the afterglow which
are used to constrain its exact location, decay rate and spectral
index (\S 2). We next discuss the properties of the host and its
environment, as observed with the {\sl Hubble Space Telescope\/} ({\sl
HST\/}) (\S 3).  We conclude by discussing the implications for the 
nature of dark bursts and for the birth places of GRBs.

     We assume a cosmology where $H_0 = 70$ km s$^{-1}$
Mpc$^{-1}$, $\Omega_m = 0.3$, and $\Omega_\Lambda = 0.7$.  In this
case a redshift of 1.0969 corresponds to a luminosity distance of 7.41
Gpc and a distance modulus of 44.35.  One arcsecond corresponds to
%17.1 comoving kpc or 
8.2 proper kpc, and the lookback time is 8.1 Gyr.

\section{The optical afterglow\label{SECTION:afterglow}}

\begin{figure}
\plotone{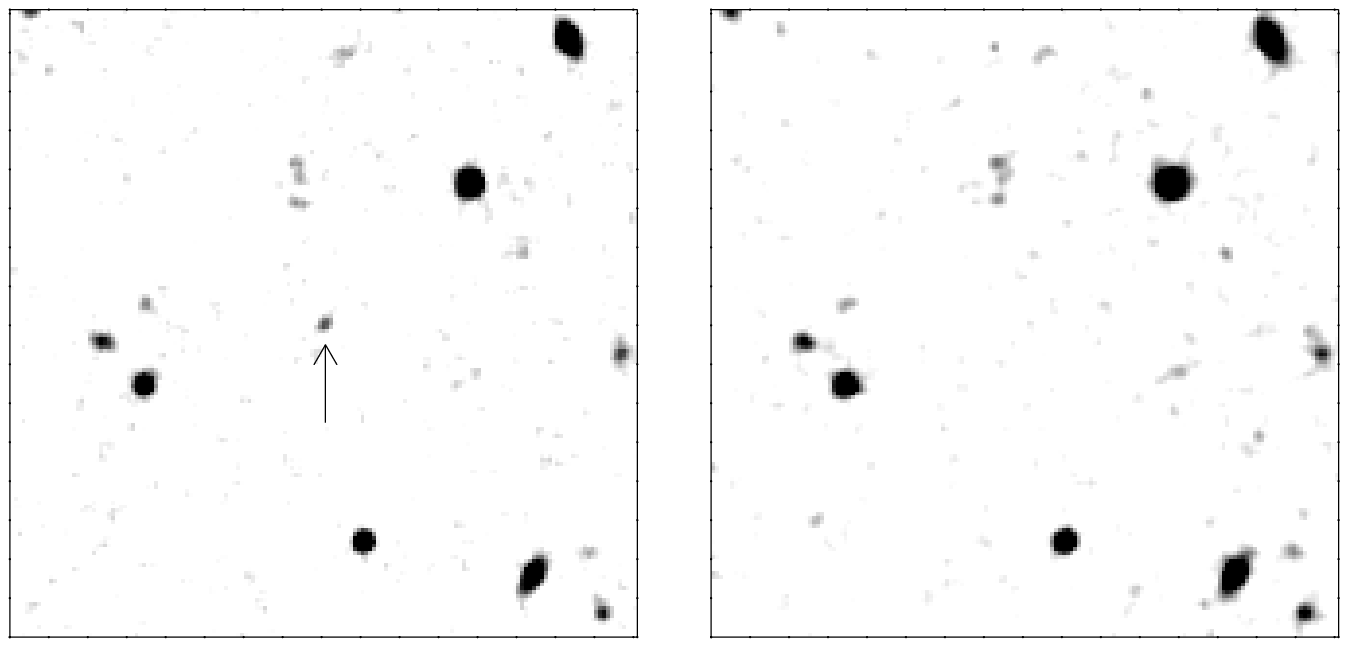}
\caption{$R$-band images
obtained with the NOT on 1998 June 13.9 UT (left) and 17.9 UT (right) which
led to the identification of the optical afterglow of
\protect\objectname{GRB 980613} marked by an arrow. The images are
$1\arcmin \times 1\arcmin$.  North is up and east is to the left.
\label{not-identification}}
\end{figure}

     The optical and near-infrared observations are reported in
Table~1. The discovery of the optical afterglow was made on the basis
of a comparison of the 1998 June 13.9 UT and 17.9 UT NOT images as described
by \citet{hjorth98} and shown in Fig.~\ref{not-identification}.
There were only two detections of the optical afterglow in each of the $R$
and $I$ bands from NOT and MDM. In the near infrared the afterglow was not
detected to a limit of $H$ = 20.9, but a faint object was however detected
with $K^\prime = 18.8\pm 0.3$. The magnitude of the host galaxy is 
$K=19.1 \pm 0.2$ (R. Chary, private communication, see also \citet{chary02}). 
Even taking the offset between $K$ and $K^\prime$
into account this indicates a substantial contribution from the host 
galaxy. For this reason we disregard the $K^\prime$ data in the remainder 
of this paper.

     \citet{djorgovski00} reported the detection of several objects
within a few arcseconds of the afterglow.  Using the {\sl HST\/} data 
presented in \S 3 we subtracted the contribution from these objects to 
derive the corrected optical afterglow magnitudes presented in Table~1. 
At discovery, 16 hours after the burst, the brightness of the optical 
afterglow was $R = 23.19 \pm 0.20$, making it one of the faintest known 
optical afterglows \citep{fynbo00c,lazzati00,bloom02x}.

     An astrometric fit to the Guide Star Catalog II yielded the afterglow 
coordinates R.A.~(J2000.0) = 10h17m57.90s, 
Dec.~(J2000.0) $= +71^{\circ}27\arcmin26\farcs3$ with an rms
uncertainty of $\pm 0\farcs27$.

%     An astrometric fit to the USNO--A2.0 catalog \citep{monet98}
%yielded the afterglow coordinates R.A.~(J2000.0) = 10h17m57.86s,
%Dec.~(J2000.0) $= +71^{\circ}27\arcmin26\farcs3$ with a random
%uncertainty of $\pm 0\farcs4$ and a systematic uncertainty of $\pm
%0\farcs25$ \citep{deutsch99}.

     As observed for most afterglows (and expected from theory cf.\ e.g.\ 
\citet{meszaros94,sari98}), the optical spectrum and light decay of the 
optical afterglow can be described by a power-law, i.e.\ 
$f_\nu \propto (t-t_{\rm GRB})^{-\alpha} \nu^{-\beta}$.  Formally, the 
decay slope derived from the $R$-band data (16--24 hours after the burst) is 
$\alpha_R=0.8\pm 0.5$ while $\alpha_I=1.3\pm 0.8$ from the $I$-band data 
(24--48 hours after the burst).  These values are consistent with the decay 
slope of $1.05\pm 0.37$ observed in the $X$-rays \citep{boer00} and with
the 1998 June 16.30 UT datapoint of \citet{djorgovski98b}.  
The formal 2$\sigma$ upper limit to the decay slope of $\alpha_R<1.8$ 
indicates that the optical afterglow did not exhibit an unusually rapid decay
during the first day after the burst. Thus, the hypothesis that rapid 
decay is the cause for the dimness of some optical afterglows 
\citep{groot98} is not applicable to \objectname{GRB 980613}.

     The spectral index was computed from the NOT $R$ and WHT $H$ data 
from 1998 June 13.9 UT, and from the MDM $R$ and $I$ data from 1998 
June 14.2.  We corrected for a Galactic extinction of $E(B\!-\!V) = 0.09$ 
\citep{SFD1998}, i.e., $A_R=0.23$, $A_I=0.16$, and $A_H=0.05$.
%We used effective wavelengths and flux conversions from \cite{fukugita95}. 
The resulting spectral indices are $\beta_\mathrm{RI} = 0.3\pm 1.1$ and
$\beta_\mathrm{RH} < 1.1$.  From the \sax\ $X$-ray flux reported by 
\cite{boer00} and \cite{soffitta01} we determine the spectral index between 
the $R$ band and the 2--10 keV $X$-ray band to be $\beta_{\rm RX}=0.59\pm0.03$, 
16 hours after the burst, consistent with $\beta_\mathrm{RI}$ and the upper 
limit on $\beta_\mathrm{RH}$. The latter limit translates into a maximum 
internal extinction in the host galaxy of 0.45 mag in $R$ in addition to the
0.23 mag of foreground extinction. 

We note that these observations are consistent with a $p=2.2$ fireball 
model ($p$ is the power-law index of the electron energy distribution)
which in the spherical adiabatic expansion (pre-break) regime \citep{sari98}
gives $\beta = 0.6$ and $\alpha = 0.9$ for $\nu < \nu_c$. However, given 
the limited amount of data available, other models are also viable.

\section{The host complex\label{SECTION:host}}

\subsection{Observations}

\begin{figure}
\plotone{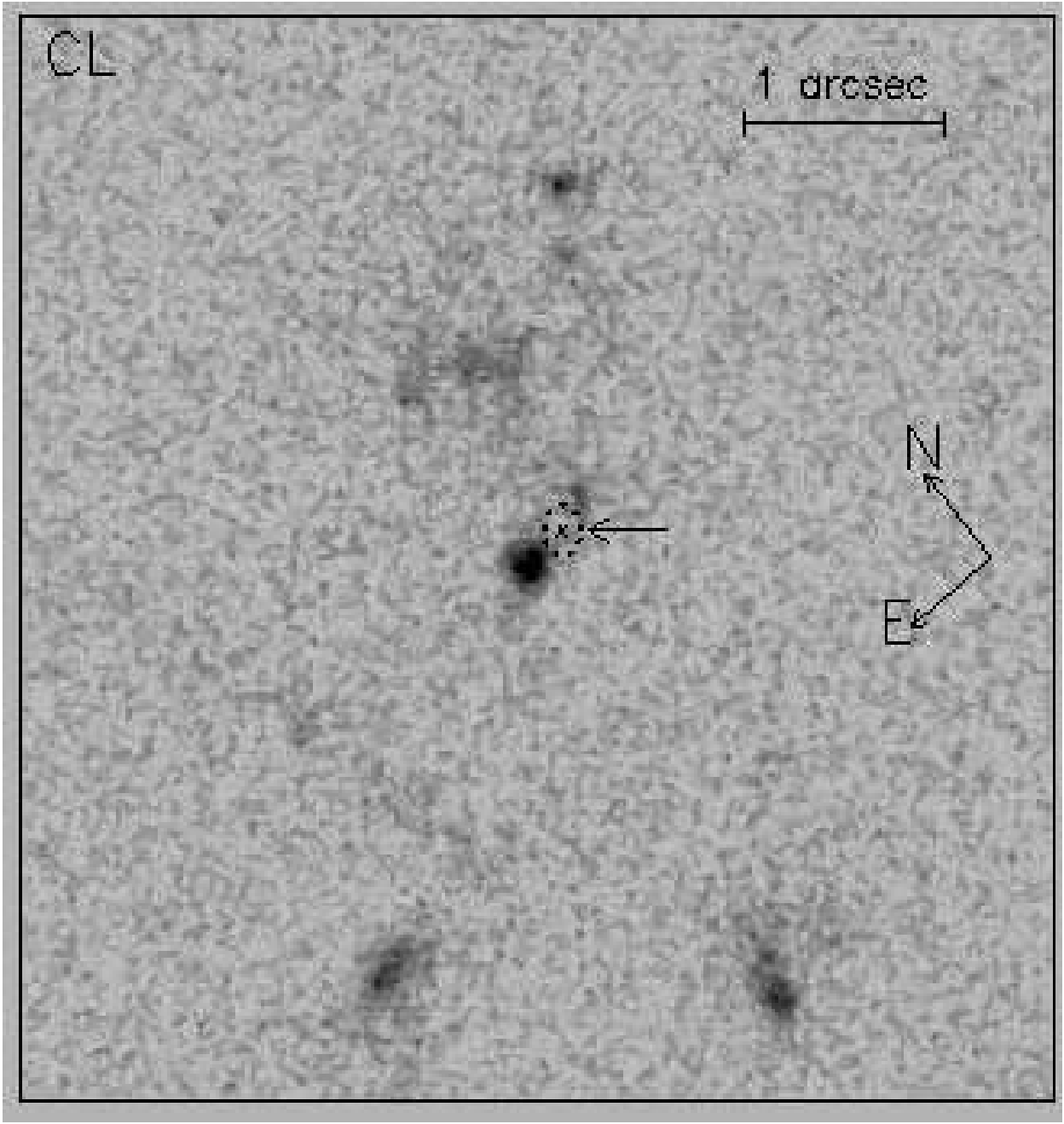}
\caption{Inverted gray-scale plot of a
$5\arcsec \times 5\arcsec$ excerpt of the {\sl HST\/}/STIS CL (50CCD
clear passband) image
of the host galaxy of \protect\objectname{GRB 980613} and its immediate
surroundings.  The location of the optical afterglow of
\protect\objectname{GRB 980613} is marked by a cross, the $2\sigma$ 
error ellipse (dashed), and an arrow. 
\label{drizzled-image}}
\end{figure}

     We used the Space Telescope Imaging Spectrograph (STIS) aboard
the {\sl HST\/} to obtain CCD images of the environment where
\objectname{GRB 980613} occurred.  These images were taken on 2000
August 20, 799 days after the burst.  The total exposure times were
5851 seconds in the 50CCD (clear; hereafter referred to as CL) aperture 
and 5936 seconds in the F28X50LP (long pass; hereafter referred to as LP)
aperture. The CCD gain was set to 1 e$^-$/ADU and the read-out noise
was 4.46 e$^-$.  We used a four-point STIS-CCD-BOX dithering pattern
with shifts of 2.5 pixels ($=0\farcs127$) between exposures. The data
were preprocessed through the standard STIS pipeline and combined using
the DITHER (v1.2) software \citep{fruchter00} as implemented in IRAF
(v2.11.3)/STSDAS (v2.1.1).  The `pixfrac' parameter was set to 0.6 and
we selected a final output scale of 0\farcs0254 pixel$^{-1}$. These
observations were taken as part of the Cycle~9 program GO-8640: A
Survey of the Host Galaxies of Gamma-Ray Bursts.  The URI for the
survey is {\tt http://www.ifa.au.dk/$\sim$hst/grb\_hosts/index.html}
\citep{holland00b}.
     In Fig.~\ref{drizzled-image} we show an inverted gray-scale plot
of a $5\arcsec \times 5\arcsec$ excerpt of the CL image, centered
on the location of \objectname{GRB 980613}. It reveals a chaotic
region consisting of numerous knots and/or interacting galaxy fragments.

\subsection{Model fitting}

     In order to obtain quantitative measures of the structural and 
photometric parameters of the objects in the host complex we made successive 
robust fits of a tilted exponential disk model convolved by the drizzled 
combination of a set of $5\times 5$ shifted and rebinned PSFs calculated by 
the TinyTim program \citep{krist95} using a subsampling of $5$. We fitted 
one object at a time, after having subtracted and/or masked all neighbouring 
objects.  The spectral type was chosen to match the 
$\mathrm{CL}-\mathrm{LP}$ color of the fitted object as closely as possible.
A telescope jitter of $0\farcs005$ was assumed.  In order to take 
proper care of charge diffusion in the STIS CCD each of the PSFs in the set 
was convolved with a $3\times 3$ Gaussian kernel appropriate for the spectral 
type used to generate the subsampled PSF.  Finally, the DITHER software was 
applied as for the observational data. 

     The model fitting is robust in the sense that the parameters are 
obtained by a maximum likelihood (ML) estimation assuming a distribution of 
the residuals which has broader wings than the normal distribution 
\citep{press92}. As an example, the median is an ML estimator if the errors 
are distributed as a symmetric two-sided exponential. We would, however, 
prefer to retain the possibility of having asymmetrically distributed
residuals. In addition, the deviation from a normal distribution should be 
small near the peak of the distribution.  The simplest probability density 
satisfying these criteria is a convolution of a Gaussian and an asymmetric 
double exponential ${1 \over 2} \gamma^{-1} (\gamma^2 - \delta^2) 
\exp( -\gamma |x| + \delta x ) $, where $\gamma > 0$ and $|\delta| < \gamma$.
The three parameters $\gamma$, $\delta$, and the Gaussian $\sigma$ are fixed 
by use of the maximum likelihood estimator on a sky annulus centered on the 
galaxy to be fitted. The residuals are scaled by the square root of the 
weights produced by the DITHER software. The robust fit of the galaxy 
parameters are obtained by seeking the minimum of minus the logarithm 
of the likelihood function, but this time with $\gamma$, $\delta$, and $\sigma$ 
held at fixed values.

\begin{figure}
\plotone{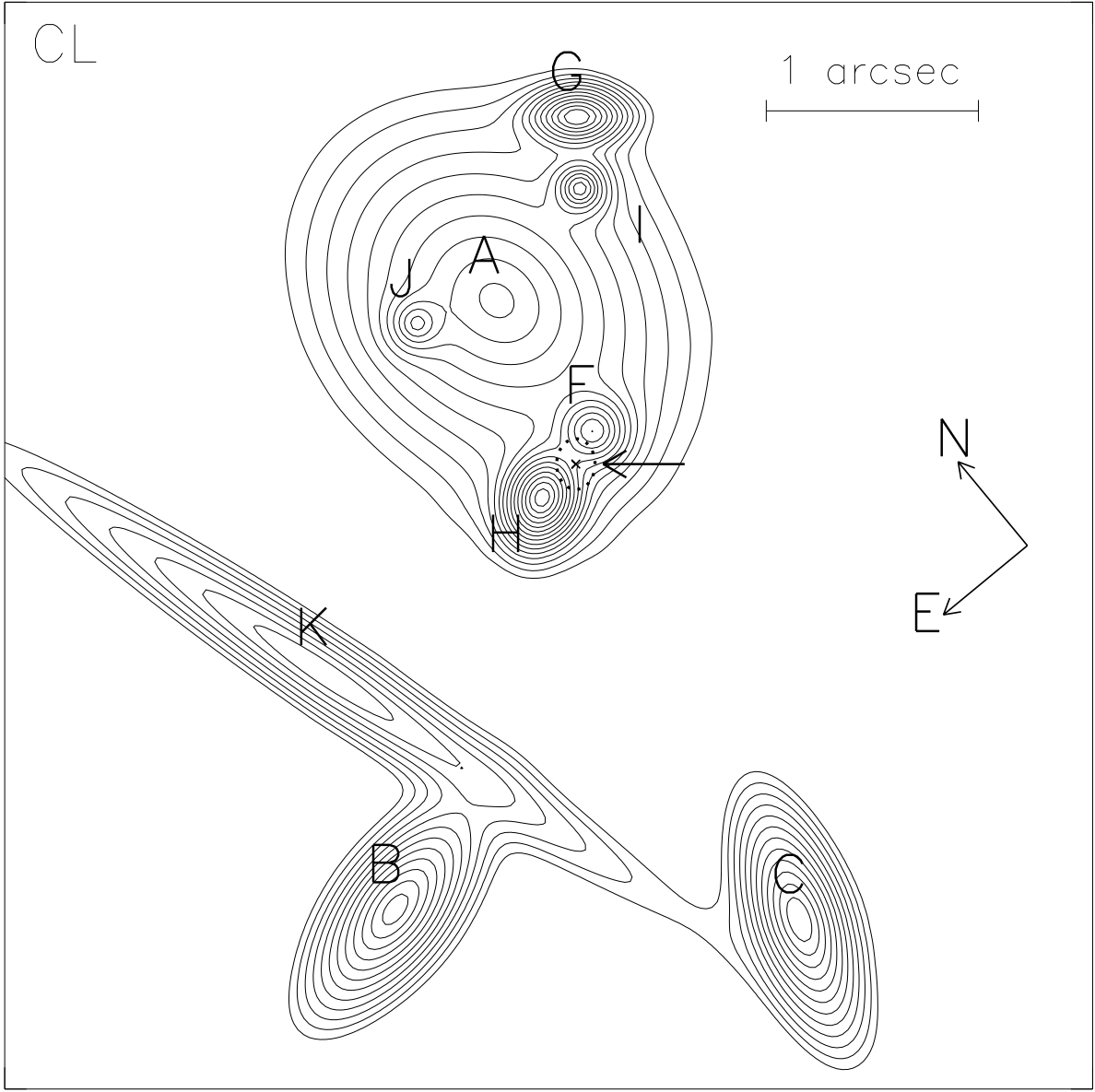}
\caption{Contour plot of a model fit to the
$5\arcsec \times 5\arcsec$ area shown in Fig.~\ref{drizzled-image}.
The location of the optical afterglow of
\protect\objectname{GRB 980613} is marked as the $2\sigma$ error ellipse
(dashed) and is marked with an arrow. A blue compact object H is evident
in the center of the image. The objects A, B, and C were identified in
ground-based images \citep{djorgovski00}. We identify F, G, H, I, J, and 
K as well. 
%North is $39\fdg54$ counterclockwise of the image y axis.
\label{contour-plot}}
\end{figure}

     The minimization is implemented in the IDL programming language by use of 
the downhill simplex method (amoeba) described by \citet{press92}.  In general 
the errors on a parameter determined by a maximum likelihood estimator are not 
symmetrically distributed around the most probable value. The error range 
corresponding to $2\sigma$ (if the method of least squares had been used) is 
defined by the parameter values that decrease the maximum of $\ln$(likelihood) 
by $2.0$ when the remaining parameters are left free to vary.  It was obtained 
by an iterative procedure. In order to obtain convergence for the smallest 
objects (F, I and J in Fig.~\ref{contour-plot}) they were forced to have 
circular symmetry.

\begin{figure}
\plotone{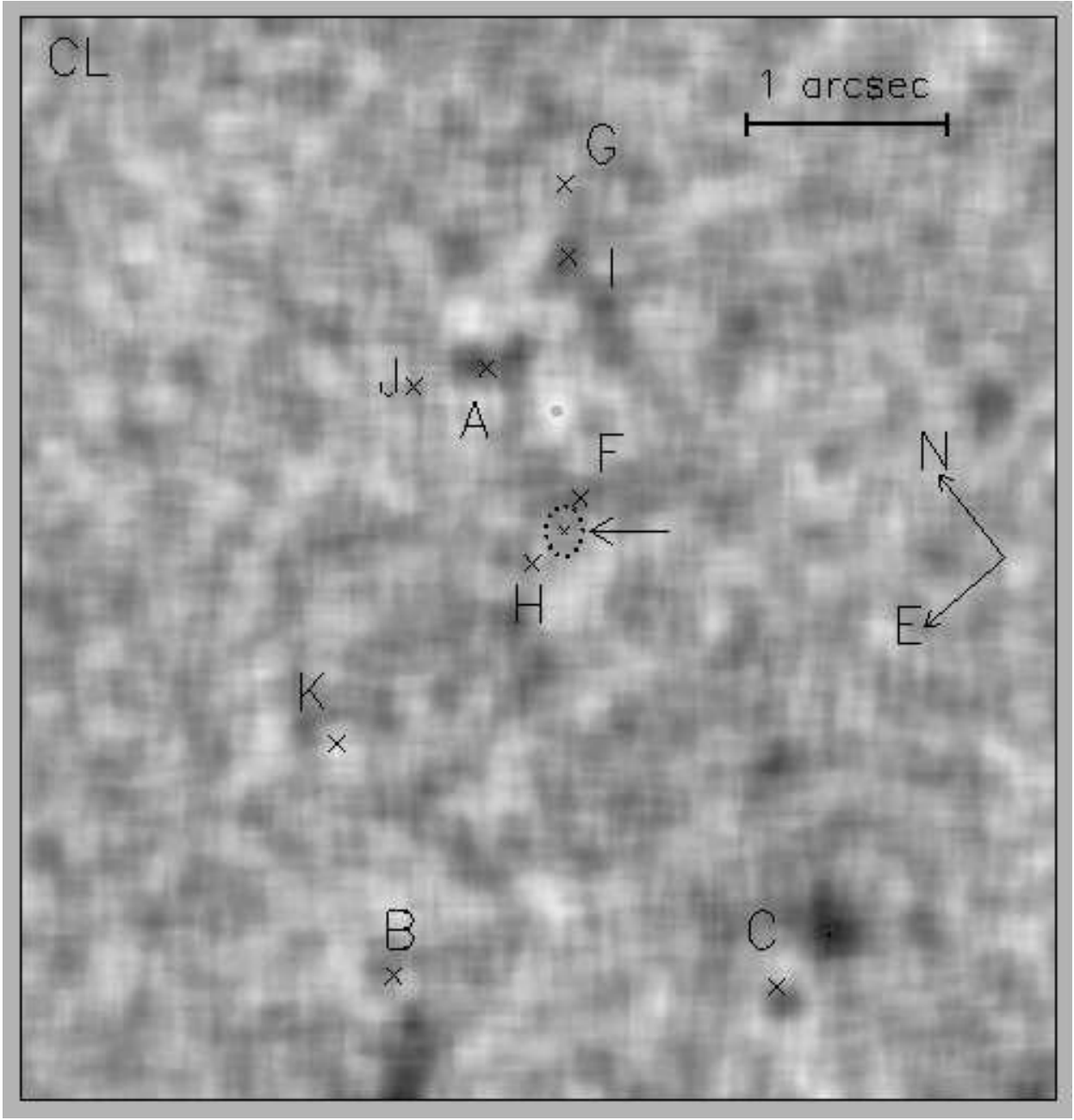}
\caption{Residual image obtained by subtracting the model
fit shown in Fig.~\ref{contour-plot} from the {\sl HST\/}/STIS CL image.
It has been smoothed by a $7 \times 7$ box in order to suppress the small 
scale noise.  The area shown is exactly the same as the one shown in 
Fig.~\ref{contour-plot}.  The crosses indicate the positions of the model 
galaxies, and the labeling is the same as in Fig.~\ref{contour-plot}.
The brightness varies from $-9$\% to $+12$\% of the average sky background.
The fit is remarkably good, except for a bar-like structure near the center
of galaxy A and some residuals close to galaxies B and C.
\label{residual-image}}
\end{figure}

Fig.~\ref{contour-plot} is a contour plot of our model fit to the image area 
shown in Fig.~\ref{drizzled-image}.  The quality of our model is demonstrated 
in Fig.~\ref{residual-image} which shows the residual image smoothed by a 
$7 \times 7$ box. The fit is remarkably good, except for a bar-like structure 
near the center of the galaxy A and some positive residuals in the outskirts 
of galaxies B and C.  In these plots we have extended the labeling introduced 
by \citet{djorgovski00}. We use A to refer to the face-on disk galaxy (to 
the north of the GRB, cf.\ \S~\ref{localization}).  We did not detect the 
very red object D nor the faint object E of \citet{djorgovski00} in either 
the CL or LP images.

The individual components have apparent diameters ranging from $0\farcs04$ 
(0.3 kpc) to $0\farcs55$ (4.5 kpc).  Their small sizes show the need for 
high-resolution ({\sl HST\/}) photometry to resolve internal details in 
GRB host galaxies and their environments.

\subsection{Localization}
\label{localization}

     We used the discovery NOT $R$-band image obtained on 1998 June
13.9 UT to determine the location of the optical afterglow in the STIS
CL image.  The host complex made a significant contribution to the 
flux in the NOT image.  Therefore we subtracted the STIS CL image
after it was transformed, scaled and smoothed to the resolution and
intensity scale of the ground-based image.  Based on four reference
sources common to the STIS CL image and the ground-based NOT image we
find \objectname{GRB 980613} to be displaced $0\farcs20\pm0\farcs05$
to the west ($\theta_0 \approx -85\degr$) of a compact, blue galaxy
(H in Fig.~\ref{contour-plot}) seen close to the center of the host
complex, towards a fainter companion (F). Both these objects are embedded 
in or projected onto the large, nearly face-on, low surface-brightness 
(the cosmological surface-brightness dimming amounts to 3.22 mag at 
$z=1.0969$) disk galaxy (A) identified by \citet{djorgovski00} as the host 
galaxy based on ground-based images. The high spatial resolution
({\sl HST\/}) images has allowed us to pinpoint the location of the GRB 
to the galaxy H which was not seen in the ground-based images of 
\citet{djorgovski00}.  We note that \cite{bloom00}, based on the
\hst\ data presented here,  also localized the GRB to the galaxy H.
As shown below, the properties of H are quite different from those of A. 

\subsection{Photometry}

We measured the color of each of the objects labeled A, B,
C, F, G, H, J, and K in Fig.~\ref{contour-plot} from robust model fits to 
the CL and LP images while keeping the structural parameters (scale length,
axial ratio, and position angle of major axis) fixed at the values obtained
from the fit of all model parameters to the CL image. 
%
%Only the position
%and the magnitude of a galaxy model were fitted. The magnitude errors are 
%nearly symmetric so the error on the color is derived as the square root 
%of the sum of squares.
For the color measurements, only the positions and magnitudes of the galaxy 
models were fitted, as it makes little sense to calculate the color 
from magnitudes obtained from different structural parameters (a procedure 
similar to using different apertures). Furthermore, for fixed structural 
parameters the magnitude errors are nearly symmetric so the error in the 
color can be derived as the square root of the sum of magnitude errors.
Note that the CL magnitudes and the structural parameters were obtained
from unrestricted fits to each galaxy in turn. The rather modest overlaps 
of some of the objects do not justify the complications of a simultaneous 
multicomponent fit (the very robust downhill simplex algorithm is not 
suitable for problems involving more than about 10 parameters).

     The STIS magnitudes CL and LP are calculated on the AB system.  We have 
used the zero points $26.386$ and $25.291$ given by \citet{gardner00} for 
1 count~s$^{-1}$ in CL and LP, respectively.  The photometric and 
morphological parameters are listed in Table~\ref{TABLE:knots}. For each 
object we present the CL total magnitude, the $\mathrm{CL}-\mathrm{LP}$ color, 
the central surface brightness $\mu_{\mathrm{CL}}(0)$ in mag~arcsec$^{-2}$, 
the scale length of the exponential disk $R_d$, the axial ratio $b/a$, and the 
position angle of the major axis. The errors given in Table~\ref{TABLE:knots} 
correspond to $2 \sigma$ errors, if the method of least squares had been used.

\subsection{Star-formation rates}

In the following we will assume that all of the objects in the host complex 
are at the same redshift---that of object A. In this case the separations 
between the objects indicate that the host complex is a compact group.  The 
best fit to the host object H has a disk scale length of $R_d = 0\farcs0360 
\pm 0\farcs0028$ ($0.30 \pm 0.02$ proper kpc).  The object is approximately 
circular with an ellipticity of $\epsilon = 1 - b/a = 0.30$. 
%The GRB occurred about midway between the host and a faint blue companion (F). 
The host object
H and its fainter companion F are the bluest objects in the field and H 
has the highest central surface brightness (Table~\ref{TABLE:knots}).

As shown in Fig.~1 of \citet{madau98} the power radiated at $\sim$2800~{\AA} 
is proportional to the instantaneous Star Formation Rate (SFR) of a model 
stellar population with a stellar birthrate of the form SFR 
$\propto \exp(-t/\tau)$, where $\tau$ is the duration of the starburst.  The 
constant of proportionality is quite insensitive to the details of the past 
star formation history. At $z = 1.0969$ the rest-frame wavelength of the 
CL bandpass is very close to 2800 {\AA}.  We used the on-line version of the 
Galaxy Isochrone Synthesis Spectral Evolution Library 
\citep{bruzual93,charlot96} and the SYNPHOT package of STSDAS under the IRAF 
environment to calculate both rest-frame and redshifted AB-magnitudes for a 
set of Spectral Energy Distributions (SEDs) for bursts of star formation 
characterized by a Salpeter IMF and an exponentially decreasing SFR with 
$\tau = 1$ Gyr. Specifically, we obtained the star-formation rate, the 
absolute magnitude, $M_B$, the apparent CL magnitude, and the 
$\mathrm{CL}-\mathrm{LP}$ color for each of 12 starbursts with solar 
metallicity at ages ranging from 1 Myr to 10 Gyr. A Galactic extinction 
of $E(B\!-\!V) = 0.09$ was applied.

\begin{figure}
\plotone{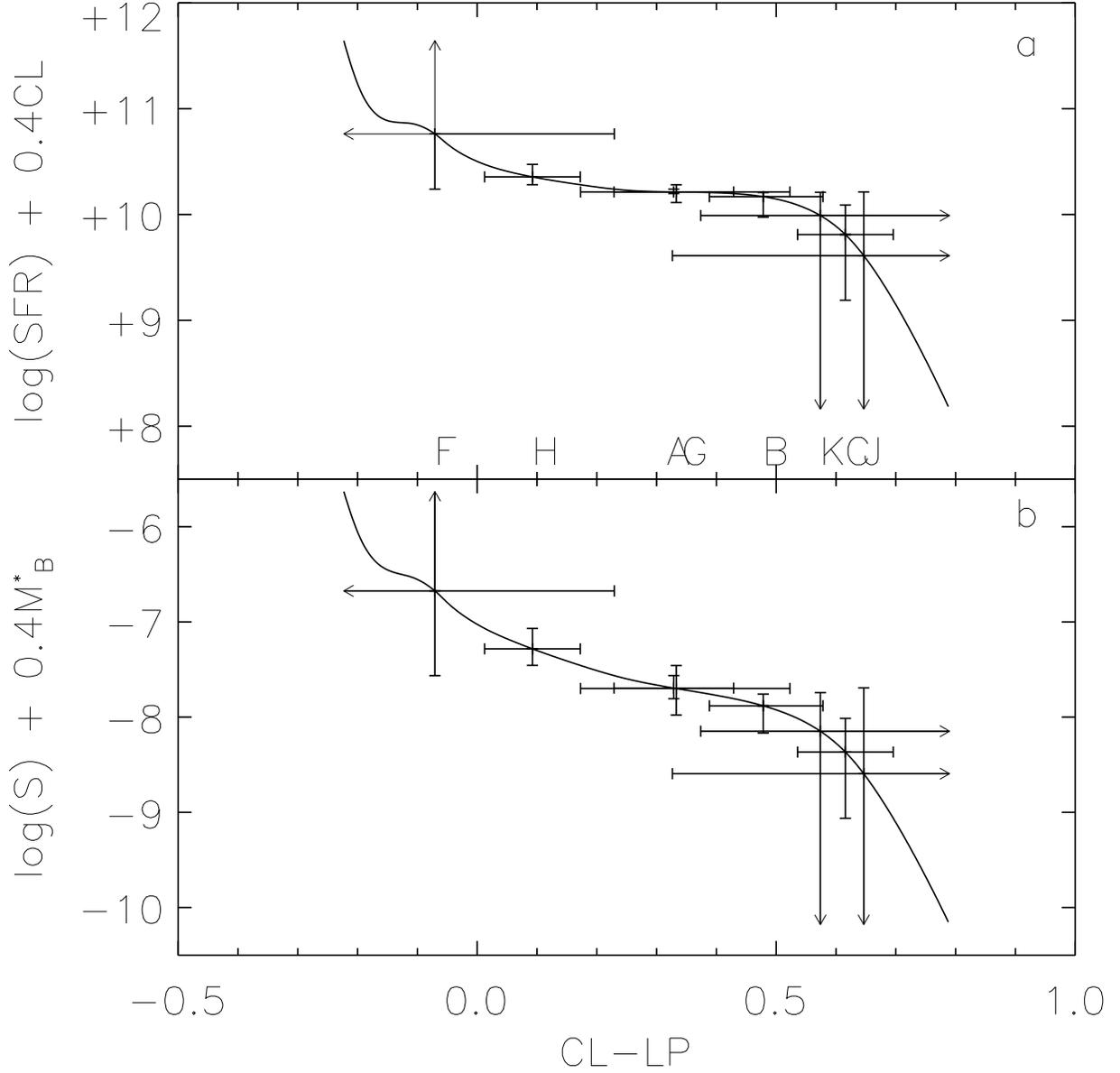}
%\caption{Specific star-formation rates of the 
%objects in the host complex in the observer's frame (a) and the rest-frame 
%(b), respectively.  Panel (a) shows $\log(\mathrm{SFR})+0.4\mathrm{CL}$ as a 
%function of the apparent color $\mathrm{CL}-\mathrm{LP}$ (CL is the
%50CCD clear passband, LP is the F28X50LP long passband.  Panel (b) shows 
%$\log(S)+0.4M^{*}_{B}$ as a function of $\mathrm{CL}-\mathrm{LP}$.
%The curves plotted represent a model with an exponentially decreasing SFR 
%with $\tau=1$~Gyr and a range of ages between 1~Myr and 10~Gyr.  SFR and 
%$S$ are in units of ${\cal M}_{\sun}\mathrm{yr}^{-1}$ and 
%${\cal M}_{\sun}\mathrm{yr}^{-1}{L^{*}_B}^{-1}$, respectively. The $2\sigma$ 
%error bars of $\mathrm{CL}-\mathrm{LP}$ are taken from Table~\ref{TABLE:knots}.
\caption{Specific star-formation rates of the objects in the host
complex expressed in the observer's frame (a) and in the rest-frame (b),
respectively.  Panel (a) shows $\log(\mathrm{SFR})+0.4\mathrm{CL}$ as a 
function of the
apparent color $\mathrm{CL}-\mathrm{LP}$ (CL is the 50CCD clear
passband, LP is the F28X50LP long passband).  
As expected, the SFR scales with the apparent magnitude CL.  Panel (b) 
shows $\log(S)+0.4M^{*}_{B}$ as a function of $\mathrm{CL}-\mathrm{LP}$. 
Note that $S$ scales with the chosen reference luminosity $L^{*}_B$. 
The curves plotted represent a model with an exponentially decreasing SFR 
with $\tau=1$~Gyr and a range of ages between 1~Myr and 10~Gyr.  SFR and 
$S$ are in units of ${\cal M}_{\sun}\mathrm{yr}^{-1}$ and
${\cal M}_{\sun}\mathrm{yr}^{-1}{L^{*}_B}^{-1}$, respectively.
The theoretical curves in (a) and (b) are used to translate the measured
$\mathrm{CL}-\mathrm{LP}$ colors with error bars into specific star-formation
rates with error bars (it is not a fit).  The $2\sigma $ error bars of 
$\mathrm{CL}-\mathrm{LP}$ are taken from Table~\ref{TABLE:knots}.
\label{specific-starformation}}
\end{figure}

     In Fig.~\ref{specific-starformation} we show the specific star-formation 
rates of the galaxies in the host complex as a function of the observed color, 
in both the observer's frame (a) and the rest frame (b). The two continuous 
curves represent a cubic spline through the 12 model values. 
 The spline curves are used to transform from color to specific SFR and
as such are also used to infer the errors.
Figures 
\ref{specific-starformation}a and \ref{specific-starformation}b show 
$\log(\mathrm{SFR})+0.4\mathrm{CL}$ and $\log(S)+0.4M^{*}_B$
($S$ is the star-formation rate per unit blue luminosity),
both as a function of the $\mathrm{CL}-\mathrm{LP}$ color.
SFR and $S$ are given in units of ${\cal M}_{\sun}{\rm yr}^{-1}$ and
${\cal M}_{\sun}{\rm yr}^{-1}{L^{*}_B}^{-1}$, respectively, where $L^{*}_B$
is the rest-frame B-band luminosity of a typical galaxy.  \citet{lilly95} find 
a typical magnitude of $M^*_B = -21.4$ (AB) for blue galaxies with 
$z \approx 1$ if $(H_0,\Omega_m,\Omega_\Lambda) = (50,1,0)$.  For our 
adopted cosmology of $(H_0,\Omega_m,\Omega_\Lambda) = (70,0.3,0.7)$ this 
corresponds to $M^*_B = -21.29$ (AB). The values in Table~\ref{TABLE:knots} 
were calculated using this value.
The $2\sigma$ error bars of $\mathrm{CL}-\mathrm{LP}$ are taken from
Table~\ref{TABLE:knots}. Note that the objects F, J, and K have
error bars extending outside the full color range of the model SEDs.
The end-points of the vertical bars are calculated as the curve values
of the end-points of the horizontal bars, except for the cases where
they extend outside the model range. These bars are limited by the
extreme values of the burst models. The specific SFR with respect to the
CL-passband (Fig.~\ref{specific-starformation}a) is, as expected, nearly
constant in the color range $0.0\lesssim\mathrm{CL}-\mathrm{LP}\lesssim0.5$.
This is, unfortunately, not the case when the SFR is expressed in terms
of the rest-frame B-band luminosity (Fig.~\ref{specific-starformation}b).

In the last two columns of Table~\ref{TABLE:knots} we present the specific 
star-formation rates in units of ${\cal M}_\sun \mathrm{yr}^{-1} {L^*_B}^{-1}$ 
for eight of the objects in the host complex, as well as $M_B-M^{*}_B$ for 
three of these objects, including the host object. The intrinsic luminosities 
for the rest of the objects are not given due to their very large 
uncertainties.  The star-formation rates listed in Table~\ref{TABLE:knots} 
and the curves shown in Fig.~\ref{specific-starformation} assume that there 
is little or no intrinsic extinction in the galaxies of the host complex.  
The star-formation rates determined from the optical images therefore 
represent lower limits to the true values.

The star-formation rates per unit blue luminosity in the compact host 
object (H) and its companion (F) are the highest of the objects listed in 
Table~\ref{TABLE:knots}.  If H is a galaxy in its own right then it is 
sub-luminous with $L_B = (0.05 \pm 0.01) L^*_B$.  However, this value is 
somewhat uncertain since $M^*_B$ is highly correlated with both the slope of 
the faint end of the galaxy luminosity function and its normalization 
\citep{lilly95}. The specific star-formation rate $S$ of the compact host is
$\approx 17$ ${\cal M}_\sun$ yr$^{-1}$ ${L^*_B}^{-1}$, the highest value of 
any known GRB host galaxy, as estimated from the 2800 \AA\ flux.  
%Table~\ref{TABLE:grbs} lists the specific star-formation rates for 
%several GRB hosts.
The star-formation rate density inside the disk scale length is
$\approx 12$ ${\cal M}_\sun$ yr$^{-1}$ kpc$^{-2}$, consistent
with values (1--1000 ${\cal M}_\sun$ yr$^{-1}$ kpc$^{-2}$)
characterizing a star burst \citep{kennicutt98}.  We stress that 
this is lower limit to the true star-formation rate because of the 
unknown UV extinction in the host galaxy (see also Chary et al.\ 2002).

\section{Discussion}

The dim optical afterglow of \objectname{GRB 980613} would have gone 
undetected under slightly less favorable search and observing conditions and, 
in this case, would have been classified as a `dark burst'
\citep{fynbo00c,lazzati00,reichart01}.  Part of the 
reason that the optical afterglow was faint may have been dust extinction 
in the host galaxy.  However, we have presented evidence that the afterglow 
suffered only at most 0.45 mag of extinction along the line of sight (at a 
redshift of 1.0969 the rest-frame reddening in the host is $A_V < 0.25$) in 
the host galaxy plus 0.23 mag of foreground extinction.  

Another explanation for the faintness of the optical afterglow may be that 
the burst was collimated but with the line of sight outside the collimated 
beam. This results in a much fainter peak afterglow magnitude while the late 
time light curve behaviour is similar to that observed if the line of sight is 
inside the collimated beam \citep{huang00}. As observed, the burst may in that 
case be expected to be weaker in $\gamma$-rays than in $X$-rays 
\citep{nakamura00}.  

The positional coincidence between \objectname{GRB 980613} and the faint 
compact blue object (H) with a high surface brightness and specific 
star-formation rate provides strong evidence in favor of the association of 
GRBs with star formation.  Not only did \objectname{GRB 980613} occur in 
a chaotic and possibly interacting region \citep{djorgovski00}, the STIS 
images presented here also reveal that the GRB occurred in what is probably 
the most active site of star formation in the entire complex.

It is interesting to note that \objectname{GRB 980613} did not
occur in the region of highest inferred total star-formation rate 
(object A) but in the region of the highest specific star-formation 
rate (objects H and F) (although the total star-formation rate of
H is also relatively high). One may speculate that GRBs trace regions 
of high specific optical star-formation rate, either because the total 
star-formation rate is in fact dramatically higher than that inferred from 
the UV flux due to dust (as in \objectname{GRB 980703}, cf.\ \cite{berger01}) 
or because the IMF could be biased towards high-mass stars in regions 
experiencing its first intensive starburst, possibly leading to a 
much higher rate of massive-star formation.  We caution however that
such conclusions should not be based on one system and note that 
\objectname{GRB 990705} \citep{andersen02} occurred in a region
of very little star-formation activity in an evolved galaxy.

\acknowledgments

     JH and AOJ thank Rolf Stabell and Sjur Refsdal for their hospitality 
at the Center for Advanced Study in Oslo where the identification of the 
optical afterglow of \objectname{GRB 980613} was made. We thank Enrico Costa 
and Eliana Palazzi for early information that facilitated the initial 
identification. This work was supported by the Danish Natural Science 
Research Council (SNF). STH would like to acknowledge support from NASA 
grant NAG5--9364. JG acknowledges the receipt of a Marie Curie
Grant from the European Commission. GB acknowledges support from the 
Icelandic Research Council and the University of Iceland Research Fund.

\clearpage

\newpage

\begin{deluxetable}{ccccccc}
%\small
%\footnotesize
%\scriptsize
%\tablewidth{7truecm}
%\tablenum{}
\tablecaption{Log of observations and photometry of the afterglow of GRB 980613
\label{tbl-1}}

\tablehead{
\colhead{Date} &
\colhead{Filter} &
\colhead{Exposure time}&
\colhead{Telescope\tablenotemark{a}}&
\colhead{FWHM} &
\colhead{Brightness\tablenotemark{b}} &
\colhead{Afterglow\tablenotemark{c}}\\
\colhead{(June 1998 UT)}&
\colhead{}&
\colhead{(sec)} &
\colhead{}&
\colhead{(arcsec)} &
\colhead{(mag)} &
\colhead{(mag)} 
}
\startdata
13.89 &  $R$        & 600           & NOT  & 0.7  & $22.81\pm0.15$ & $23.19\pm0.20$\\
13.89 &  $K^\prime$ & $5\times 450$ & CAHA & 1.2  & $18.8\pm0.3$   &     \\
13.90 &  $H$        & 2400          & WHT  & 1.2  & $>20.9$        & \\
14.20 &  $I$        & $6\times 500$ & MDM  & 0.96 & $22.64\pm0.08$ & $23.13\pm0.20$\\
14.24 &  $R$        & $4\times 500$ & MDM  & 1.14 & $22.96\pm0.09$ & $23.53\pm0.20$\\
14.90 &  $K^\prime$ & $4\times 600$ & CAHA & 1.2  & $>18.5$        & \\
15.19 &  $I$        & $4\times 900$ & MDM  & 0.99 & $23.25\pm0.15$ & $24.1\pm0.4$\\
17.91 &  $R$        & 600           & NOT  & 0.68 & $>24.4$        &  \\
\enddata

\tablenotetext{a}{
NOT is the 2.56-m Nordic Optical Telescope on La Palma, equipped with
ALFOSC (1998 June 13.89) and HiRAC (1998 June 17.89); CAHA is the
3.5-m telescope at Calar Alto, equipped with OMEGA-Cass; WHT is the
4.2-m Willian Herschel Telescope on La Palma, equipped with CIRSI; MDM
is the 2.4-m Hiltner telescope of the MDM Observatory on Kitt Peak,
equipped with a thinned, back-side illuminated SITe CCD.}
\tablenotetext{b}{
The $R$ and $I$ data were calibrated to the Cousins system using
\citet{landolt92} standard stars \citep{halpern98}. The quoted
magnitudes refer to a 1\arcsec\ radius aperture.}
\tablenotetext{c}{
The $R$ and $I$ magnitudes after subtracting scaled and transformed
{\sl HST\/}/STIS images to correct for the contribution from the host
environment.}

\end{deluxetable}

%%%%%%%%%%%%%%%%%%%%%%%%%%%%%%%%%%%%%%%%%%%%%%%%%%%%%%%%%%%%%%%%%%%%%%%%%%%

%\begin{deluxetable}{cccc}
%%\small
%%\footnotesize
%%\scriptsize
%\tablewidth{10.0truecm}
%%\tablenum{}
%\tablecaption{Photometry of host complex 
%\label{TABLE:photometry}}
%\tablehead{
%\colhead{Object} &
%\colhead{CL} &
%\colhead{$\mathrm{CL} - \mathrm{LP}$} &
%\colhead{$\mu_{\mathrm{CL}}(0)$} \\
%\colhead{} &
%\colhead{(mag)} &
%\colhead{(mag)} &
%\colhead{(mag arcsec$^{-2}$)}
%}
%\startdata
%A       & $24.82$  $^{+0.10}_{-0.11}$ & $ 0.33 \pm 0.10$ & $24.02$ \\
%B       & $25.80$  $^{+0.09}_{-0.09}$ & $ 0.48 \pm 0.10$ & $22.62$ \\
%C       & $25.83$  $^{+0.09}_{-0.09}$ & $ 0.62 \pm 0.08$ & $22.43$ \\
%F       & $27.34$  $^{+0.20}_{-0.21}$ & $-0.07 \pm 0.30$ & $22.39$ \\
%G       & $27.11$  $^{+0.14}_{-0.14}$ & $ 0.33 \pm 0.19$ & $21.44$ \\
%H       & $26.08$  $^{+0.05}_{-0.05}$ & $ 0.09 \pm 0.08$ & $20.86$ \\
%I       & $28.10$  $^{+0.27}_{-0.27}$ &       ------     & $21.55$ \\
%J       & $28.06$  $^{+0.25}_{-0.27}$ & $ 0.65 \pm 0.41$ & $22.28$ \\
%K       & $26.03$  $^{+0.19}_{-0.20}$ & $ 0.57 \pm 0.23$ & $24.40$ \\
%\enddata
%\end{deluxetable}

%%%%%%%%%%test %%%%%%%%%%%%%%%%%%%%%%%%%%%%%%%%%%%%%%%%%%%%%%%%%%%%%%%%%%%%%%%%%

\begin{deluxetable}{ccccccrcc}
%\small
%\footnotesize
%\scriptsize
\tabletypesize{\scriptsize}
%\tablewidth{15.0truecm}
%\tablenum{}
\tablecaption{Photometry, morphology and star-formation rates of the
host complex
\label{TABLE:knots}}
\tablehead{
\colhead{Object} &
\colhead{CL} &
\colhead{$\mathrm{CL} - \mathrm{LP}$} &
\colhead{$\mu_{\mathrm{CL}}(0)$} &
\colhead{R$_d$} &
\colhead{$b/a$} &
\colhead{P.A.} &
\colhead{$\log(S)$} &
\colhead{$M_B-M^{*}_B$} \\
\colhead{} &
\colhead{(mag)} &
\colhead{(mag)} &
\colhead{(mag arcsec$^{-2}$)} &
\colhead{(mas)} &
\colhead{} &
\colhead{($\degr$)} &
\colhead{$({\cal M}_\sun \mathrm{yr}^{-1} {L^*_B}^{-1})$} &
\colhead{}
}
\startdata
A       & $24.82$  $^{+0.10}_{-0.11}$ & $ 0.33 \pm 0.10$ & $24.02$ 
& $ 277$ $^{+26}_{-24}$   & $0.80$ & $  6$ 
& $0.82^{+0.13}_{-0.11}$ & $1.3 \pm 0.3$ \\
B       & $25.80$  $^{+0.09}_{-0.09}$ & $ 0.48 \pm 0.10$ & $22.62$ 
& $91.9$ $^{+9.9}_{-9.1}$ & $0.39$ & $104$ 
& $0.64^{+0.12}_{-0.29}$ &     \nodata  \\
C       & $25.83$  $^{+0.09}_{-0.09}$ & $ 0.62 \pm 0.08$ & $22.43$
& $83.0$ $^{+9.8}_{-9.0}$ & $0.34$ & $160$ 
& $0.15^{+0.36}_{-0.69}$ &     \nodata  \\
F       & $27.34$  $^{+0.20}_{-0.21}$ & $-0.07 \pm 0.30$ & $22.39$
& $41.0$ $^{+13}_{-12}$   & $1.00$ &  \nodata 
& $ 1.8^{+1.0 }_{-0.9 }$ &     \nodata   \\
G       & $27.11$  $^{+0.14}_{-0.14}$ & $ 0.33 \pm 0.19$ & $21.44$ 
& $29.3$ $^{+9.5}_{-9.5}$ & $0.31$ & $ 52$
& $0.82^{+0.24}_{-0.28}$ & $3.6 \pm 0.5$ \\
H       & $26.08$  $^{+0.05}_{-0.05}$ & $ 0.09 \pm 0.08$ & $20.86$ 
& $36.0$ $^{+2.8}_{-2.8}$ & $0.60$ & $122$
& $1.23^{+0.21}_{-0.17}$ & $3.3 \pm 0.3$ \\
I       & $28.10$  $^{+0.27}_{-0.27}$ &       \nodata     & $21.55$ 
& $19.5$ $^{+15}_{-12}$   & $1.00$ &  \nodata 
& \nodata  &     \nodata \\
J       & $28.06$  $^{+0.25}_{-0.27}$ & $ 0.65 \pm 0.41$ & $22.28$ 
& $27.9$ $^{+13}_{-10}$   & $1.00$ &  \nodata  
& $-0.1^{+0.9 }_{-1.6 }$ &     \nodata \\
K       & $26.03$  $^{+0.19}_{-0.20}$ & $ 0.57 \pm 0.23$ & $24.40$ 
& $ 188$ $^{+40}_{-35}$   & $0.07$ & $ 16$ 
& $ 0.4^{+0.4 }_{-2.0 }$ &     \nodata \\
\enddata
\tablecomments{The meaning of the quoted errors is discussed in
\S\S\ 3.2 and 3.4}
\end{deluxetable}

\end{document}